\documentstyle[aps,twocolumn,prl]{revtex}

\begin{document}
\draft
\preprint{}
\title{
Superconducting condensation energy and neutron resonance in high-$T_{c}$ superconductors
}
\author{
Jian-Xin Li}
\address{
National Laboratory of Solid States of Microstructure and Department of Physics, Nanjing University, Nanjing 210093, China}

\author{Chung-Yu Mou}
\address{
Department of Physics, National Tsing Hua University, Hsinchu 30043,
Taiwan}

\author{Chang-De Gong}
\address{
Chinese Center of Advanced Science and Technology (World Laboratory), P.O. Box 8730,\\
Beijing 100080, China\\
and National Laboratory of Solid States of Microstructure, Nanjing University,
Nanjing 210093, China}

\author{T. K. Lee}
\address{
Institute of Physics, Academia Sinica, Taipei 11529, Taiwan\\
and National Center for Theoretical Sciences, P.O.Box 2-131, Hsinchu, Taiwan}
\maketitle

\begin{abstract}
Based on the slave-boson theory of the two-dimensional $t-t'-J$ model, we calculate the superconducting condensation energy for 
optimally doped and overdoped high $T_{c}$ cuprates at finite temperatures
using a renormalized random phase approximation. The contributions come from the
mean-field part and the antiferromagnetic spin fluctuations.
In the presence of neutron resonance peak at $(\pi,\pi)$, the latter 
is shown to have similar temperature and doping dependences as the difference in antiferromagnetic (AF) exchange energy between the normal and the superconducting state.
This difference has been proposed to be related to the superconducting condensation energy by Scalapino and White. 
The total condensation energy, however, is about 1/2 smaller than the proposed AF exchange energy difference and
shows a more rapid decrease as the temperature rises. The doping dependence of the condensation
energy is found to be consistent with experiments. In particular, near zero temperature, our result shows a good quantitative agreement with experiments.
\end{abstract}
\pacs{PACS number: 74.25.-q,74.25.Bt,75.40.Gb,74.72.-h}

\section{INTRODUCTION}

One of the important aspects to understand the mechanism of superconductivity
is to investigate the origin of the condensation energy,
which is the free energy difference
between the normal and the superconducting(SC) states. In conventional BCS superconductors,
the condensation energy is accounted for by the change in the ion kinetic
energy between the normal and the SC phases, as shown by Chester~\cite{che}.
In high-$T_{c}$ superconductors,
it is now widely accepted that the strong electronic
interactions rather than the electron-phonon interactions may be responsible for
the superconductivity.  However, no consensus has been achieved 
on the detail pairing
mechanism, consequently on the origin of
the condensation energy~\cite{leg,hirsh,scala}.

In analogy with the phonon-mediated electron pairing,
Scalapino and White~\cite{scala}
proposed that if the pairing is mediated by the antiferromagnetic (AF) exchange
interaction, the condensation energy
will be proportional to the change in the AF exchange energy between the
normal and the SC states. This gives a direct connection
between the condensation
energy and the dynamical spin susceptibility $\chi$ (or the spin structure factor $S({\bf q},
\omega)={\rm Im}\chi ({\bf q},\omega)/[1+\exp(-\omega/T)]$) as~\cite{scala,demler,aba},
\begin{eqnarray}
E_{N}-E_{S}&=&{3\over 2}J\sum_{q}\int_{0}^{\infty} {d\omega \over \pi}[{\rm Im}
\chi^{N}({\bf q},\omega)-{\rm Im}\chi^{S}({\bf q},\omega)]  \nonumber \\
& & \times [\cos(q_{x})+\cos(q_{y})]{1 \over {1-\exp(-\omega/T)}}. \label{scalapino}
\end{eqnarray}
The dynamical spin susceptibility can be probed directly by neutron scattering experiments.
The most prominent feature in these experiments when the temperature decreases below the SC transition is the appearance
of a sharp resonance peak at the AF wavevector ${\bf Q}=(\pi,\pi)$ ($\pi$ resonance) in the SC state~\cite{ros,moo,fong}.
From this observation, Demler and Zhang~\cite{demler} further argued that the
appearance of the neutron resonance will cause the antiferromagnetic exchange energy
to be lower than that in the normal state. Without resorting to any microscopic models,
they estimated the exchange energy to be around 0.016$J$ based on the experimental data for the spin
structure factor and the $q$-width of the resonance peak. The estimated condensation energy is roughly of the same
magnitude as that obtained by experiments~\cite{loram} and therefore accounts for a large part of the condensation energy.
The above estimate is based on Eq.(\ref{scalapino}) without including the kinetic energy.
Furthermore, it is limited to zero temperature and with the assumption that the spin susceptibility in the normal
state has negligible contribution to Eq.(1). Its validity thus relies on further tests against
experimental data. In particular, as pointed out at the end of Ref.~\cite{demler},
a useful test of this idea is to compare the theoretical estimate with experiments
at finite temperatures. Experimentally, because the net difference of the
AF exchange energy is small, it will require extremely careful measurements to check Eq.(\ref{scalapino}).
Only recently, a study of the temperature dependence is performed. In Ref.~\cite{dai}, the neutron scattering measurement shows a similar temperature dependence between the condensation energy
from the AF exchange interactions and the resonance peak intensity for optimal doped YBa$_{2}$Cu$_{3}$O$_{6+x}$.
In view of the experimental situation, it would be useful to carry out calculations
at finite temperatures based on an existing high-$T_{c}$ model. Recently, the $t-t^{\prime}-J$ model has been shown 
to produce the resonance peak~\cite{bri,li2}, hence it would be interesting to verify its prediction for the condensation energy.

In this paper, we examine the temperature and doping dependences of
the condensation energy in a microscopic model -- the two-dimensional (2D)
$t-t^{\prime}-J$ model. To avoid the complications due to the appearance of the pseudo-gap in the normal state, 
we shall confine our calculations in the optimal and overdoped regions\cite{note1}.
Theoretically, the condensation energy can be calculated from the difference in the thermodynamic
potential between the normal and the SC states because the difference in free energy in the SC state is equal to the difference in thermodynamical potential~\cite{ric}. In order to include the $\pi$ resonance peak, we calculate the thermodynamical potential in the linked cluster expansion via a renormalized
random-phase approximation (RPA) as defined in Ref.~\cite{bri}. This approach has been previously tested in different contexts~\cite{bri,li2,Mou,Mou1} and has been shown to capture many important features of the $\pi$ resonance peak.
Here we shall further investigate how it would predict for the temperature dependence of the condensation energy.
We show that the condensation energy coming from
the AF exchange correction is nearly the same as Eq.(\ref{scalapino}) due to the presence of the neutron resonance peak.
Furthermore, the condensation energy from the AF exchange correction has the similar temerature dependence as the resonance peak intensity for both optimally doped and overdoped systems. This is consistent with the
experimental results observed in the optimally doped sample~\cite{dai}.
However, the total condensation energy after including the mean-field free energy difference
between the normal and SC states exhibits a more rapid drop in its temperature dependence than that calculated with Eq.(\ref{scalapino}). We also investigate the doping dependence of the total condensation energy and find it in agreement
with the experiment. In view of the appearance of the pseudo-gap in the normal state for
underdoped cuprates~\cite{tim} where a part of the condensation energy has
gained above $T_{c}$~\cite{loram}, this result indicates that the condensation energy for the optimally doped and overdoped cuprates should show more rapid temperature drop than that for the underdoped cuprates.

This paper is organized as follows. In Sec.II, we introduce the model and define notations. The condensation energy is
derived using the linked cluster expansion. In Sec.III, we calculate the temperature and doping dependences of the condensation energy and compare them with that calculated based on Eq.(\ref{scalapino}).
Finally, we give a concluding remark in Sec.IV.

\section{THERMODYNAMIC POTENTIAL AND CONDENSATION ENERGY}

We start with the 2D $t-t^{\prime}-J$ model which reads,
\begin{eqnarray}
H&=&-\sum_{<ij>,\sigma}tc^{\dag}_{i\sigma}c_{j\sigma}-h.c.
-\sum_{<ij>',\sigma}t'c^{\dag}_{i\sigma}c_{j\sigma}-h.c. \nonumber \\
& & +J \sum_{<ij>}{\bf S}_{i} \cdot {\bf S}_{j},
\end{eqnarray}
where, $<ij>$ denotes the nearest-neighbor (n.n.) bond, $<ij>'$ the next n.n. bond and ${\bf S}_{i}={1\over 2}c_{i\alpha}^{\dagger}\sigma_{\alpha \beta}c_{i\beta}$. In the slave-boson method, the physical electron operators $c_{i\sigma}$ are expressed by slave bosons $b_{i}$ carrying the charge and fermions
$f_{i\sigma}$ representing the spin; $c_{i\sigma}=b_{i}^{+}f_{i\sigma}$. In the SC state,
we consider the order parameters $\Delta_{ij}=<f_{i\uparrow}f_{j\downarrow}-f_{i\downarrow}f_{j\uparrow}>=\pm\Delta_{0}$ with the $d$-wave symmetry and $\chi_{ij}=\sum_{\sigma} <f_{i\sigma}^{+}f_{j\sigma}>=\chi_{0}$, 
in which bosons condense $b_{i}\rightarrow <b_{i}>=\sqrt \delta$ ($\delta$ is the hole concentration)~\cite{ubb}.
Then, the mean-field Hamiltonian in the SC state is,
\begin{eqnarray}
H_m&=&\sum_{k\sigma}\epsilon_{k}f^{\dag}_{k\sigma}f_{k\sigma}
-\sum_{k}\Delta_{k} (f^{\dag}_{k\uparrow}f^{\dag}_{-k\downarrow}+h.c.) \nonumber \\
& & +2NJ'(\chi_{0}^{2}+\Delta_{0}^{2}), \label{Hm}
\end{eqnarray}
where $\epsilon_{k}=-2(\delta t+J'\chi_{0})[\cos(k_{x})+\cos(k_{y})]-
4\delta t'\cos(k_{x})\cos(k_{y})-\mu$ is the dispersion for fermions,
and $\Delta_{k}=2J'\Delta_{0}[\cos(k_{x})-\cos(k_{y})]$, with
$J'=3J/8$.
In the optimal and overdoped regions, the mean-field Hamiltonian for the normal state is
obtained by setting the SC gap $\Delta_{0}=0$~\cite{fuk}.
The mean-field parameters $\chi_{0}$, $\Delta_{0}$ ($\Delta_{0}$=0 in
the normal state) and the chemical potential $\mu$ for different
doping $\delta$ are obtained from a self-consistent calculation~\cite{ubb}.

It has been shown~\cite{bri,li2,fuk} that the spin susceptibility at the mean-field level represented by the first fermionic bubble in Fig.1 fails to describe some important features observed by inelastic neutron scattering
experiments, such as the resonance peak~\cite{ros,moo,fong}. The underlying reason is because
the mean-field theory underestimates the spin fluctuation near $(\pi,\pi)$\cite{lee2}.
Therefore, it is necessary to go beyond the mean field approximation.
Formally, this can be done by perturbing around the mean-field Hamiltonian, i.e., we write 
$H=H_m+H^{\prime }$, and treat $H^{\prime }$ as a perturbation. In principle, all the fluctuations are included.
However, different selection of subset diagrams may result in different kinds of fluctuations.
For the spin fluctuation, the usual random phase approximation selects a series of ring diagrams as 
shown in Fig.1. The resulting spin susceptibility can be resummed as
\begin{equation}
\chi^{+-}({\bf q},\omega)={ \chi_{0}^{+-}({\bf q},\omega) \over {1+\eta J\gamma_{q}
\chi_{0}^{+-}({\bf q},\omega)}}. \label{RPA}
\end{equation}
where $\gamma_{q}=\cos (q_{x})+\cos (q_{y})$, $\chi_{0}$ is the unperturbed spin susceptibility which comes from the fermionic bubble. The parameter $\eta$ is formally introduced to keep track of the renormalization of vertex and its value is one in the usual RPA approach. However, the spin susceptibility calculated using Eq.(\ref{RPA})
with $\eta=1$ leads to an AF instability for doping $\delta\le \delta_{c}\approx 0.22$, which is much larger than the experimental data around $\delta_{c}=0.02$. It indicates that there exists other fluctuations which acts as to suppress this overestimation. In the renormalized RPA approach\cite{bri}, one models the suppression of $\eta$  by treating $\eta$ as a phenomenological parameter, whose value is determined by setting the AF instability at the observed value $\delta=0.02$.
For the material parameters we use, we found $\eta=0.34$ for the SC state and $\eta=0.15$ for the normal state.
Using $\eta=0.34$ for the SC state, Brinckmann and Lee have explained the resonance peak and the incommensuration structure observed in neutron scatterings, and some of the present authors have quantitatively accounted for the doping
dependences of the peak/dip/hump structure in the angle-resolved
photoemission spectra\cite{li2} and the tunneling spectra in a NS junction\cite{Mou1}.

The above considerations can be carried over to the calculation of the thermodynamic potential. We consider the same 
set of ring diagrams in the linked cluster expansion of the thermodynamic potential as shown in Fig.2, and find it is 
given by~\cite{mahan}
\begin{equation}
\Delta \Omega ={1 \over \beta}\sum_{i\omega_{n}}\sum_{q}\int_{0}^{1}
d\nu {J\gamma_{q}\chi_{0}^{+-}({\bf q},i\omega_{n})
\over {1+\nu \eta J\gamma_{q}\chi_{0}^{+-}({\bf q},i\omega_{n})}}
\end{equation}
Here, we have used the renormalized strength $\eta J$ for the vertex. The parameter $\nu$ is used to keep track of the number of times that the potential appears in the perturbation expansion. Since $\nu$ enters in the same way as a coupling constant, the integration can be taken outside of the summation over $i\omega_{n}$.  Performing the summation over $i\omega_{n}$, we get
\begin{equation}
\Delta \Omega =J\sum_{q}\int_{0}^{1}d\nu \int_{0}^{\infty}
{d\omega \over \pi}{\rm Im}\chi^{+-}(\nu,{\bf q},\omega)\gamma_{q}
\coth ({\beta \omega \over 2})
\end{equation}
with
\begin{equation}
\chi^{+-}(\nu,{\bf q},\omega)={ \chi_{0}^{+-}({\bf q},\omega) \over {1+\nu \eta J\gamma_{q}
\chi_{0}^{+-}({\bf q},\omega)}} \label{chi}
\end{equation}
In obtaining Eq.(\ref{chi}), we have made use of the symmetry ${\rm Im}\chi ({\bf q},-\omega)=-{\rm Im}\chi ({\bf q},\omega)$.

The condensation energy $H_{c}^{2}v/8\pi$ per unit cell is equal to the difference of the thermodynamic potential(free energy) between the normal and the SC states. In our approach, it includes the contribution from the AF exchange correction $\Delta \Omega^{\prime}$ and that from the mean-field Hamiltonian. The free energy at the mean-field
level in the SC state can be easily calculated from Eq.(\ref{Hm}) as,
\begin{eqnarray}
F_{0}^{S}&=&-{2\over \beta}\sum_{k}\ln(1+e^{-\beta E_{k}})+\sum_{k}(\epsilon_{k}-E_{k}) \nonumber \\ 
& & +2NJ^{\prime}(\Delta_{0}^{2}+\chi_{0}^{2})
\end{eqnarray}
with $E_{k}=\sqrt {\epsilon_{k}^{2}+\Delta_{k}^{2}}$. Its counterpart in the 
normal state $F_{0}^{N}$ can be obtained from Eq.(8) by setting $\Delta_{0}=0$. 
To include the contribution from $\chi$, we note that the mean-field free energy already contains the Hartree-Fock contribution, i.e., the one-loop term in Fig.2\cite{schrieffer}. Therefore, we should substract it to avoid double counting
and arrive at,
\begin{eqnarray}
& &{H_{c}^{2}v\over 8\pi}=  \nonumber \\
& &{3J \over2}\sum_{q}\int_{0}^{1}d\nu \int_{0}^{\infty}
{d\omega \over \pi}[{\rm Im}\chi^{N}(\nu,{\bf q},\omega)-
{\rm Im}\chi^{S}(\nu,{\bf q},\omega)]  \nonumber \\
& & \times \gamma_{q} \coth ({\omega \over 2T})-{3J \over 2}\sum_{q}\int_{0}^{\infty}
{d\omega \over \pi}[{\rm Im}\chi_{0}^{N}({\bf q},\omega)-{\rm Im}\chi_{0}^{S}({\bf q},\omega)] \nonumber \\
& & \times \gamma_{q}\coth ({\omega \over 2T})
+F_{0}^{N}-F_{0}^{S}. \label{ourcondensation}
\end{eqnarray}
where, we have abbreviated $\chi^{+-}$ as $\chi$. The first term of the 
right hand side of Eq.(9) comes from the AF fluctuation correction as shown in Fig.2, while the second term is the Hartree-Fock contribution which should be substracted from the first term. Finally, $F_{0}^{N}-F_{0}^{S}$ is the free energy difference at the mean-field level.

Before performing the numerical calculation, we would like to point out that all the relevant energies such as
the AF exchange energy, the kinetic energy and the energy for the formation 
of spin gap are contained in Eq.(\ref{ourcondensation}).
Comparing the AF exchange energy in Eq.(\ref{ourcondensation}) to that in Eq.(\ref{scalapino}), 
we find two differences, a) there is an additional integral over $\nu$ in Eq.(\ref{ourcondensation}) and 
b) the temperature factors $\coth (\omega /2T)$ and $1/[1-\exp(-\omega/T)]$ are different. At zero temperature or if the main contribution to the integral over $\omega$ is dominated by those satisfying $\omega/T \gg 1$,
the temperature factors are identical. When $\nu=1$, Eq.(\ref{chi}) is the same as the renormalized
spin susceptibility Eq.(\ref{RPA}). Therefore, if the integral over $\nu$ is dominated by that part
around $\nu \approx 1$ and the integral over $\omega$ is dominated by the spectral weight
around $\omega/T \gg 1$, Eq.(\ref{ourcondensation}) can be reduced to Eq.(\ref{scalapino}). We will discuss this issue in
detail in the first part of the following section.

\section{NUMERICAL CALCULATIONS}

In this section, we carry out the calculations of the condensation energy based on Eq.(\ref{ourcondensation}) and 
Eq.(\ref{scalapino}). The purpose of this section is twofold.
First, we will show that when a resonance peak occurs in the spin excitation spectrum, the result obtained by 
Scalapino and White Eq.(\ref{scalapino}) is almost identical to the condensation energy calculated
from the perturbation diagrams shown in Fig.2. Second, we will investigate the temperature and doping dependences 
of the condensation energy based on Eq.(\ref{ourcondensation}), and compare them with the experiments.

Numerical calculations are performed by dividing the Brillouin zone into $128\times 128$ lattices with $t=2J$ and $t'=-0.45t$ as usual. A quasi-particle damping $\Gamma=0.02J$ is used.
We find that the results for the condensation energy (including its magnitudes) nearly
does not change when $\Gamma$ is changed around $0.02J$, though the spin susceptibility
indeed follows the change of $\Gamma$. The reason is because the change of $\Gamma$ induces nearly identical changes 
for the spin susceptibility in both the normal and the SC states.

The temperature dependences of the difference in the AF exchange energy calculated with Eq.(\ref{scalapino})
are presented in Fig.3 for doping densities $\delta$=0.16, 0.18 and 0.20, respectively. The results for the
first term of the right hand side of Eq.(\ref{ourcondensation}) with the same parameters are
shown in Fig.4. In all cases, both results are only weakly $T$-dependent
at low temperatures, but fall off rapidly as $T$ is increased towards $T=0.08J$.
Comparing Figs.3 and 4, one can see that the overall temperature
dependences for both cases are remarkably similar for the doping range investigated.
Meanwhile, both have a similar magnitude, though the latter is slightly smaller than
the former. Thus, we may conclude that the first term of Eq.(\ref{ourcondensation}) and Eq.(\ref{scalapino}) may describe the same physics. Therefore, it provides a thermodynamical extension to the work by Scalapino and White\cite{scala}.
As noted above, the most possible reason for the same temperature dependence is that the integral over $\nu$ is dominated by
that part around $\nu=1$ and the integral over frequency is dominated by that part satisfying
$\omega/T \gg 1$. To see the variation of the integrand in Eq.(\ref{ourcondensation}) with $\nu$, we show in Fig.5 the results for Im$\chi^{+-}(\nu,{\bf Q},\omega)$, which is given by 
Eq.(\ref{chi}), at temperature $T=0.005t$ and doping $\delta=0.16$ for several values of $\nu$ (1, 0.9 and 0.5). When $\nu=1$, the integrand $\chi^{+-}(\nu,{\bf q},\omega)$ is identical to the spin susceptibility Eq.(\ref{RPA}).
The most remarkable change in the spin susceptibility across the SC transition is the
emergence of the resonance peak at momentum ${\bf Q}=(\pi,\pi)$ below $T_{c}$, which is shown
as the solid line in Fig.5. In the framework of the $d$-wave BCS theory, the origin of the neutron resonance peak has been attributed to a collective spin excitation mode~\cite{bri,li2,blu,li1}, which corresponds
to $1+\eta J\gamma_{Q}{\rm Re}\chi^{+-}_{0}({\bf Q},\omega)=0$ and Im$\chi^{+-}_{0}({\bf Q},\omega)$ approaches to zero.
It is caused by the step-like rise of the imaginary part of the unperturbed spin susceptibility Im$\chi_{0}$ at its threshold as shown in the inset of Fig.5, where the solid line denotes its real part, 
the dashed line its imaginary part and the dotted line represents $-1/\nu\eta J\gamma_{Q}$ with $\nu=1$.
According to the Kramers-Kroenig relation, a logarithmic singularity in its real part Re$\chi_{0}$ occurs due to this step-like rise, which exhibits as a peak in the numerical calculations on finite lattices. This enhancement in Re$\chi_{0}$ shifts downward the position of the resonance mode, which is the cross between the dotted line and the solid line in the inset of Fig.5, into the SC gap where no damping for spin excitations is expected.
When $\nu$ decreases, the dotted line in the inset of Fig.5 will rise, so the position of the cross will move to higher frequencies above the threshold where the dampings become nonzero. As $\nu$ decreases further, there is no cross anymore. However, if $1+\eta J\gamma_{Q}{\rm Re}\chi_{0}({\bf Q},\omega)$ is small, a peak still occurs. But the intensity of
the peak will decrease, such as the dashed line for $\nu=0.9$ in Fig.5. This may be regarded as the proximity effect of the resonance mode. As $\nu$ moves far away from 1, no peak exists and only a broad hump appears such as the dotted line in
Fig.5 for $\nu=0.5$.  From the figure, one can see that the integrated spectral weight over $\omega$ in the SC state decreases with the decrease of $\nu$. However, little change occurs in its normal state results (the extrapolated value). Therefore, it is the existence of the resonance peak that the integral over $\nu$ is dominated by that part around $\nu=1$. Because the resonance peak is around $\omega=0.52J$( 0.53$J$ and 0.52$J$ for $\delta=0.18$ and
$\delta=0.20$, respectively), so the temperature factors $\coth (\omega/T)$
and $1/[1-\exp(-\omega/T)]$ are essentially equal to 1 even for the highest
temperature $T=0.08J$ considered here. As for the momentum distribution,
we wish to point out two facts: a) the resonance peak appears at $\bf Q=(\pi,\pi)$ and
drops rapidly as ${\bf q}$ moves away from $\bf Q$; b) the summation over $q$
is weighted by a factor $\gamma_{q}=\cos(q_{x})+\cos(q_{y})$ being
maximum at $\bf Q$. Obviously, both help to filter out the contributions around the resonance peak.
Therefore, the very similar temperature dependence of the results calculated with
Eq.(\ref{scalapino}) and with the first term of Eq.(\ref{ourcondensation}) suggests that a large part of the
condensation energy due to the AF exchange interactions comes from the spectral weight
around the resonance mode, as first argued by Demler and Zhang\cite{demler}.

Because the integral over $\nu$ in Eq.(\ref{ourcondensation}) is dominated by the contributions around $\nu=1$ and only in this range of $\nu$ is the integrand Im$\chi^{+-}(\nu,{\bf Q},\omega)$ approximately identical to the spin susceptibility Eq.(\ref{RPA}), it leads the magnitudes of the condensation energy calculated with the first term of Eq.(\ref{ourcondensation}) to be smaller than those calculated with Eq.(\ref{scalapino}), as
one can see from a comparison of Figs.3 and 4. In Fig.4, we also show the temperature dependences of the resonance peak intensities for doping $\delta=$0.16(squares), 0.18(circles) and 0.20(triangles). For all doping levels from the optimal
to overdoped regimes, we find that the peak intensity and the condensation
energy due to the AF exchange interactions follow the same temperature dependence.
This result is in good agreement with the recent experiment on YBa$_{2}$Cu$_{3}$O$_{6.93}$~\cite{dai}
which is an optimally doped system, though the result for the overdoped
system waits for future experiment testing. This coincidence further supports the above conclusion that the contribution to the condensation energy from the AF exchange corrections arises mainly from
spin excitations around the resonance peak.

The total condensation energy  based on the renormalized RPA approach is plotted as a function of temperature in Fig.6 for
different dopings $\delta=0.16, 0.18$ and 0.20. From a comparison with Fig.3,
we find that the results calculated here exhibit different temperature dependence, i.e.,
they decrease more rapidly with the rise of temperatures. This difference is due to the rapid variation with temperature of the contribution from the mean-field part. Meanwhile, their magnitudes are only about half of those calculated with Eq.(\ref{scalapino}). We note that the condensation energy for doping
$\delta=0.16$(nearly optimally doped) at zero temperature is around $0.023J$, which is in reasonably agreement with the estimated value $0.016J$\cite{demler} based on the experimental data for the spin structure factor and the $q$-width of 
the resonance peak in YBa$_{2}$Cu$_{3}$O$_{7}$~\cite{fong} and also with the experimental data 
$E_{c}=6$K($0.23J\approx 30$K if $J=1300$K is taken) in the specific-heat measurement\cite{loram}. Thus, our investigation based on the renormalized RPA approach to the $t-t^{\prime}-J$ model gives a quantitative account
for the condensation energy. 

The doping dependence of the condensation energy calculated with Eq.(\ref{ourcondensation}) is shown in Fig.7. 
An obvious feature is that the condensation energy decreases as the doping density rises from the optimally doped to the overdoped regime. This is related to the doping dependence of the maximum SC gap $\Delta_{0}$ at the mean-field level which shows the same trend, for example $\Delta_{0}$=0.335J, 0.322J and 0.311J for $\delta=0.16, 0.18$ and 0.2 at $T=0.005J$, respectively. Moreover, an approximate linear variation of the condensation energy with
doping is found. This result is consistent with the experimental data determined from the heat capacity measurements\cite{tallon}. However, we note that the condensation energies shown in Fig.6 do not approach to zero
when the temperature is as high as $0.08J$ which may correspond to 104K if
we choose $J=1300$K. This is the fault of the slave-boson mean-field calculation which over-estimates $T_c$ so that $\Delta_{0}$ does not approach to zero even at $T=0.08J$. Therefore, 
even though our approach gives good quantitative results near zero temperature,
it may break down at temperatures near true transition temperatures.

\section{CONCLUDING REMARKS}

In summary, based on the slave-boson theory of the
two-dimensional $t-t'-J$ model, we calculate the superconducting condensation energy for 
optimally doped and overdoped high-$T_{c}$ cuprates at finite temperatures
using a renormalized random phase approximation. 
It is composed of the contributions from the mean-field Hamiltonian and
that from the antiferromagnetic spin fluctuation.

We show that the contribution from the antiferromagnetic spin fluctuation is
essentially the same as the argument by Scalapino and White\cite{scala}, which relates the condensation energy to the difference of the antiferromagnetic exchange energy between the normal state and the superconducting state,
due to the existance of the neutron resonance mode. However, both the temperature
dependence and the magnitudes of the total condensation energy after including the
mean-field part are different from those due to the AF exchange energy.
Our result for the total condensation energy gives a quantitative account for the observed
condensation energy near zero temperature. We also calculate the doping dependence of the condensation energy
and find it in agreement with experiments.

Taking into account of our previous studies of the relationship of
the resonance neutron peak to the angle-resolved photoemission spectra
\cite{li2} and the tunneling spectra\cite{Mou1}, we believe that the resonant spin collective mode
plays an important role in determining many physical properties in the superconducting state of
high-$T_{c}$ cuprates. A particular interesting conclusion from this work is that the 
renormalized random phase approximation seems to have already captured the main features 
of the resonance mode. However, it is introduced phenomenologically and its derivation
is left for future work.

\section*{ACKNOWLEDGMENTS}

JXL was supported by the National Nature Science Foundation of China. CYM
thanks the support from NSC of Taiwan under Grant
No. NSC 89-2112-M-007-091 and TKL for NSC 89-2112-M-001-103.
CDG thanks the support by the Ministry of Science and Technology of China(NKBRSF-G19990646).

\newpage

\section*{FIGURE CAPTIONS}
Fig.1 Feynman diagram for the renormalized spin susceptibility calculated in the
random phase approximation(RPA). The solid lines represent the fermionic Green's functions and the dashed line the antiferromagnetic coupling constant $J\gamma_{q}$.

\vspace{0.3cm}
Fig.2 Feynman diagram for the contribution to the thermodynamic potential from antiferromagnetic fluctuations. The solid lines represent the fermionic Green's functions and the dashed line
the antiferromagnetic coupling constant $J\gamma_{q}$.

\vspace{0.3cm}
Fig.3 Temperature dependences of the difference in antiferromagnetic exchange energy
between the normal and the superconducting states Eq.(\ref{scalapino}) for hole concentrations
$\delta=$0.16, 0.18 and 0.20, which is argued by Scalapino and white~\cite{scala}
to be the condensation energy.

\vspace{0.3cm}
Fig.4 Temperature dependences of the condensation energy due to the antiferromagnetic
fluctuation correction [the first term in the right hand of Eq.(9)] for hole concentrations $\delta=$0.16, 0.18 and 0.20. The solid symbols denote the intensities of the resonance peakes at different temperatures (squares for $\delta=0.16$, circles for $\delta=0.18$ and
triangles for $\delta=0.20$). The intensity is scaled in order to compare it with the condensation energy.

\vspace{0.3cm}
Fig.5 Frequency dependences of Im$\chi^{+ -}(\nu,{\bf Q},\omega)$ at temperature $T=0.005J$,
${\bf Q}=(\pi,\pi)$ and doping $\delta=0.16$ for $\nu=$1, 0.9 and 0.5, respectively.
Inset shows the bare spin susceptibility Im$\chi_{0}({\bf Q},\omega)$.
The solid line denotes its real part and the dashed line its imaginary part.
The dotted line represents $-1/\eta J\gamma_{\bf Q}$(see text). Similar results are obtained
for other dopings expected that the positions of the resonance peak are shifted, so are not
shown here.

\vspace{0.3cm}
Fig.6 Temperature dependences of the total condensation energy calculated with Eq.(\ref{ourcondensation}) for hole concentrations $\delta=$0.16, 0.18 and 0.20. They show more rapid drops in comparison to Fig.3.

\vspace{0.3cm}
Fig.7 Doping dependence of the total condensation energy calculated with Eq.(\ref{ourcondensation}) at $T=0.005J$.


\begin{thebibliography}{99}
\bibitem{note1} This was also the regime suggested by Scalapino and White~\cite{scala}, that could be used to check the validity of Eq.(\ref{scalapino}).
\end{thebibliography}
\end{document}